\documentclass[
 reprint,
 amsmath,amssymb,
 aps,
 ]{revtex4-2}

\usepackage{graphicx}
\usepackage{dcolumn}
\usepackage{bm}
\usepackage[pdfstartview=FitH,
CJKbookmarks=true,
bookmarksnumbered=true,
bookmarksopen=true,
colorlinks,
linkcolor=blue,
anchorcolor=blue,
citecolor=blue,
allcolors=blue,
pdfborder=001,
]{hyperref}
\usepackage[caption=false]{subfig}
\usepackage{floatrow}
\usepackage{enumitem}
\DeclareSubrefFormat{myparens}{#1~(#2)}
\DeclareCaptionListOfFormat{myparens}{#1(#2)}
\usepackage{xspace}
\usepackage{multirow}
\usepackage{makecell}
\usepackage{placeins}

\RequirePackage[normalem]{ulem} 
\RequirePackage{color}\definecolor{RED}{rgb}{1,0,0}\definecolor{BLUE}{rgb}{0,0,1}

\providecommand{\DIFaddbegin}{} 
\providecommand{\DIFaddend}{} 
\providecommand{\DIFdelbegin}{} 
\providecommand{\DIFdelend}{}

\providecommand{\DIFaddbeginFL}{} 
\providecommand{\DIFaddendFL}{} 
\providecommand{\DIFdelbeginFL}{} 
\providecommand{\DIFdelendFL}{}

\newcommand{\DIFscaledelfig}{0.5}
\RequirePackage{settobox} 
\RequirePackage{letltxmacro} 
\newsavebox{\DIFdelgraphicsbox} 
\newlength{\DIFdelgraphicswidth} 
\newlength{\DIFdelgraphicsheight} 
\LetLtxMacro{\DIFOincludegraphics}{\includegraphics} 
\newcommand{\DIFaddincludegraphics}[2][]{{\color{blue}\fbox{\DIFOincludegraphics[#1]{#2}}}} 
\newcommand{\DIFdelincludegraphics}[2][]{
\sbox{\DIFdelgraphicsbox}{\DIFOincludegraphics[#1]{#2}}
\settoboxwidth{\DIFdelgraphicswidth}{\DIFdelgraphicsbox} 
\settoboxtotalheight{\DIFdelgraphicsheight}{\DIFdelgraphicsbox} 
\scalebox{\DIFscaledelfig}{
\parbox[b]{\DIFdelgraphicswidth}{\usebox{\DIFdelgraphicsbox}\\[-\baselineskip] \rule{\DIFdelgraphicswidth}{0em}}\llap{\resizebox{\DIFdelgraphicswidth}{\DIFdelgraphicsheight}{
\setlength{\unitlength}{\DIFdelgraphicswidth}
\begin{picture}(1,1)
\thicklines\linethickness{2pt} 
{\color[rgb]{1,0,0}\put(0,0){\framebox(1,1){}}}
{\color[rgb]{1,0,0}\put(0,0){\line( 1,1){1}}}
{\color[rgb]{1,0,0}\put(0,1){\line(1,-1){1}}}
\end{picture}
}\hspace*{3pt}}} 
} 
\LetLtxMacro{\DIFOaddbegin}{\DIFaddbegin} 
\LetLtxMacro{\DIFOaddend}{\DIFaddend} 
\LetLtxMacro{\DIFOdelbegin}{\DIFdelbegin} 
\LetLtxMacro{\DIFOdelend}{\DIFdelend} 
\DeclareRobustCommand{\DIFaddbegin}{\DIFOaddbegin \let\includegraphics\DIFaddincludegraphics} 
\DeclareRobustCommand{\DIFaddend}{\DIFOaddend \let\includegraphics\DIFOincludegraphics} 
\DeclareRobustCommand{\DIFdelbegin}{\DIFOdelbegin \let\includegraphics\DIFdelincludegraphics} 
\DeclareRobustCommand{\DIFdelend}{\DIFOaddend \let\includegraphics\DIFOincludegraphics} 
\LetLtxMacro{\DIFOaddbeginFL}{\DIFaddbeginFL} 
\LetLtxMacro{\DIFOaddendFL}{\DIFaddendFL} 
\LetLtxMacro{\DIFOdelbeginFL}{\DIFdelbeginFL} 
\LetLtxMacro{\DIFOdelendFL}{\DIFdelendFL} 
\DeclareRobustCommand{\DIFaddbeginFL}{\DIFOaddbeginFL \let\includegraphics\DIFaddincludegraphics} 
\DeclareRobustCommand{\DIFaddendFL}{\DIFOaddendFL \let\includegraphics\DIFOincludegraphics} 
\DeclareRobustCommand{\DIFdelbeginFL}{\DIFOdelbeginFL \let\includegraphics\DIFdelincludegraphics} 
\DeclareRobustCommand{\DIFdelendFL}{\DIFOaddendFL \let\includegraphics\DIFOincludegraphics} 
\makeatletter 
\let\sout@orig\sout 
\renewcommand{\sout}[1]{\ifmmode\text{\sout@orig{\ensuremath{#1}}}\else\sout@orig{#1}\fi} 
\makeatother 
\RequirePackage{listings} 
\RequirePackage{color} 
\lstdefinelanguage{DIFcode}{ 
  moredelim=[il][\color{red}\sout]{\%DIF\ <\ }, 
  moredelim=[il][\color{blue}\uwave]{\%DIF\ >\ } 
} 
\lstdefinestyle{DIFverbatimstyle}{ 
	language=DIFcode, 
	basicstyle=\ttfamily, 
	columns=fullflexible, 
	keepspaces=true 
} 
\lstnewenvironment{DIFverbatim}{\lstset{style=DIFverbatimstyle}}{} 
\lstnewenvironment{DIFverbatim*}{\lstset{style=DIFverbatimstyle,showspaces=true}}{} 
\begin{document}

\preprint{APS/123-QED}

\newcommand{\BESIIIorcid}[1]{\href{https://orcid.org/#1}{\hspace*{0.1em}\raisebox{-0.45ex}{\includegraphics[width=1em]{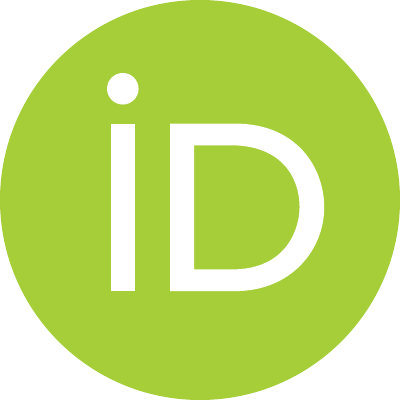}}}} 

\newcommand\MeV{\ensuremath{\mathrm{MeV}}}
\newcommand\GeV{\ensuremath{\mathrm{GeV}}}

\setlength{\abovedisplayskip}{6pt}
\setlength{\belowdisplayskip}{6pt}

\title{\boldmath Observation of the $X(2370)$ in $J/\psi\rightarrow\gamma K^{0}_{S}K^{0}_{S}\pi^{0}$ and $J/\psi\rightarrow\gamma \pi^{0}\pi^{0}\eta$
}

\author{
\begin{small}
\begin{center}
M.~Ablikim$^{1}$\BESIIIorcid{0000-0002-3935-619X},
M.~N.~Achasov$^{4,c}$\BESIIIorcid{0000-0002-9400-8622},
P.~Adlarson$^{83}$\BESIIIorcid{0000-0001-6280-3851},
X.~C.~Ai$^{89}$\BESIIIorcid{0000-0003-3856-2415},
C.~S.~Akondi$^{31A,31B}$\BESIIIorcid{0000-0001-6303-5217},
R.~Aliberti$^{39}$\BESIIIorcid{0000-0003-3500-4012},
A.~Amoroso$^{82A,82C}$\BESIIIorcid{0000-0002-3095-8610},
Q.~An$^{78,65,\dagger}$,
Y.~H.~An$^{89}$\BESIIIorcid{0009-0008-3419-0849},
M.~S.~Anderson$^{39}$\BESIIIorcid{0009-0008-1550-2632}
% ...
\\
\vspace{0.2cm}
(BESIII Collaboration)\\
\vspace{0.2cm} {\it
$^{1}$ Institute of High Energy Physics, Beijing 100049, People's Republic of China\\
% ...
$^{4}$ Budker Institute of Nuclear Physics SB RAS (BINP), Novosibirsk 630090, Russia\\
% ...
$^{31}$ INFN Sezione di Ferrara, (A)INFN Sezione di Ferrara, I-44122, Ferrara, Italy; (B)University of Ferrara, I-44122, Ferrara, Italy\\
% ...
$^{39}$ Johannes Gutenberg University of Mainz, Johann-Joachim-Becher-Weg 45, D-55099 Mainz, Germany\\
% ...
$^{65}$ State Key Laboratory of Particle Detection and Electronics, Beijing 100049, Hefei 230026, People's Republic of China\\
% ...
$^{78}$ University of Science and Technology of China, Hefei 230026, People's Republic of China\\
% ...
$^{82}$ University of Turin and INFN, (A)University of Turin, I-10125, Turin, Italy; (B)University of Eastern Piedmont, I-15121, Alessandria, Italy; (C)INFN, I-10125, Turin, Italy\\
$^{83}$ Uppsala University, Box 516, SE-75120 Uppsala, Sweden\\
% ...
$^{89}$ Zhengzhou University, Zhengzhou 450001, People's Republic of China\\
\vspace{0.2cm}
$^{\dagger}$ Deceased\\
% ...
$^{c}$ Also at the Novosibirsk State University, Novosibirsk, 630090, Russia\\
% ...
}
\end{center}
\vspace{0.4cm}
\end{small}
}

{
\renewcommand{\baselinestretch}{1.20}

\begin{abstract} 

Based on $(10087\pm44)\times10^{6}$ $J/\psi$ events collected with the BESIII detector, the $J/\psi\rightarrow\gamma K^{0}_{S}K^{0}_{S}\pi^{0}$ and $J/\psi\rightarrow\gamma \pi^{0}\pi^{0}\eta$ processes are studied. The $X(2370)$ is observed in both the $K^{0}_{S}K^{0}_{S}\pi^{0}$ and $\pi^{0}\pi^{0}\eta$ invariant mass spectra, with statistical significances greater than $14\sigma$ and $20\sigma$, respectively. By combining measurements from these processes with those from the previously reported $J/\psi\rightarrow\gamma K^{0}_{S}K^{0}_{S}\eta^{\prime}$ process, the mass and width of the $X(2370)$ are determined to be $2359^{+13}_{-14}~\text{MeV}/c^{2}$ and $170^{+44}_{-29}~\text{MeV}$, respectively. In addition, the decay $X(2370)\to a_{0}(980)^{0}\pi^{0}$ with $a_{0}(980)^{0}\to \pi^{0}\eta$ is observed with a statistical significance exceeding $9\sigma$. 
The properties of the $X(2370)$\textemdash decay pattern similarities to that of $\eta_{c}$, are consistent with those of a pseudoscalar glueball.
\end{abstract}
\vspace{-2.0mm}
}
\maketitle

Quantum Chromodynamics (QCD), a non-Abelian SU(3) gauge theory, predicts the existence of glueballs, which are uniquely formed by self-interacting gluons \textemdash the gauge bosons of the strong interactions. The search for glueballs in experiments is crucial for testing and developing QCD. Predictions of the $0^{-+}$ glueball mass from different lattice QCD (LQCD) groups range from 2.3 to 3.0~$\GeV/c^{2}$~\cite{LQCD1,LQCD2,LQCD3,LQCD4,LQCD5,Vadacchino:2023vnc}. However, although $J/\psi$ radiative decays are considered an ideal place for glueball searches~\cite{Kopke:1988cs,PhysRevD.50.3268,PhysRevD.55.5749}, no $0^{-+}$ glueball candidates had been observed in $J/\psi$ radiative decays in the mass range above $2.3~\GeV/c^{2}$ until 2011.

In 2011, based on a sample of 0.2 billion $J/\psi$ events collected with the BESIII detector, the $X(2370)$ particle was first observed in the process $J/\psi\rightarrow\gamma \pi^{+}\pi^{-}\eta^{\prime}$ with a statistical significance greater than $6.4\sigma$. Its mass and width were measured to be $2376.3 \pm 8.7(\text{stat})^{+3.2}_{-4.3} (\text{syst})~\MeV/c^{2}$ and $83\pm17(\text{stat})^{+44}_{-6}(\text{syst})~\MeV$~\cite{1835_confirmed}, respectively. The observation is particularly interesting because the measured mass is consistent with the range predicted by LQCD for the lightest pseudoscalar glueball. Moreover, $J/\psi$ radiative decays are generally considered a favorable environment for glueball production, and a pseudoscalar glueball is expected to decay into channels such as $\pi^{+}\pi^{-}\eta^{\prime}$~\cite{1835_confirmed}. In subsequent studies, using 1.3 billion $J/\psi$ events, the $X(2370)$ was confirmed in the decays $J/\psi\rightarrow \gamma K^{+}K^{-}\eta^{\prime}$ and $J/\psi\rightarrow \gamma K^{0}_{S}K^{0}_{S}\eta^{\prime}$ with a statistical significance of $8.3\sigma$~\cite{epjc.s10052_gongli}. Furthermore, using 10 billion $J/\psi$ events, the spin-parity quantum numbers of the $X(2370)$ were determined to be $0^{-+}$ with a significance greater than $9.8\sigma$ in the $J/\psi\rightarrow\gamma K^{0}_{S}K^{0}_{S}\eta^{\prime}$ process~\cite{2370jpc}. Its mass and width were measured to be $2395 \pm 11 ({\rm stat})^{+26}_{-94}({\rm syst})~\MeV/c^{2}$ and $188^{+18}_{-17}({\rm stat})^{+124}_{-33}({\rm syst})~\MeV$, respectively. The product branching fraction of $J/\psi\to\gamma X(2370)\to \gamma f_{0}(980)\eta^{\prime} \to \gamma K^{0}_{S}K^{0}_{S} \eta^{\prime}$ was measured to be $\left( 1.31 \pm 0.22 ({\rm stat})^{+2.85}_{-0.84}({\rm syst}) \right) \times 10^{-5}$. The measured mass of the pseudoscalar $X(2370)$ is within the range predicted by LQCD for the lightest pseudoscalar glueball~\cite{LQCD5,Vadacchino:2023vnc}. In addition to the glueball interpretation, other interpretations and phenomenological studies of the $X(2370)$ have also been proposed~\cite{Sun:2021kka,Yu:2011ta,Dong:2020okt,Wang:2025nme,Su:2022eun}. To understand the nature of the $X(2370)$, more decay modes should be searched for and studied using 10 billion $J/\psi$ decays collected at BESIII.

\begin{figure}[htbp]
\centering
\vspace{-5mm}
\subfloat{
  \includegraphics[width=0.48\textwidth]{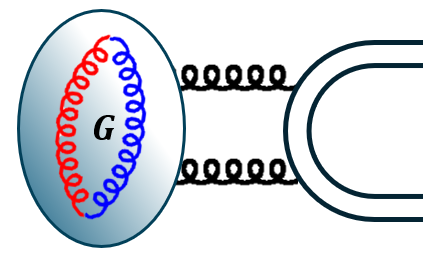}
  \label{fig:XDecay_glueball}
\put(-124, 57){(a)}
}
\hspace{-2mm}
\subfloat{
  \includegraphics[width=0.48\textwidth]{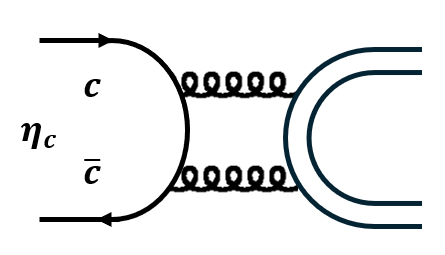}
  \label{fig:XDecay_etac}
\put(-124, 57){(b)}
}
\caption{Decays of (a) the $0^{-+}$ glueball and (b) the $\eta_{c}$.}
\label{fig:XDecay}
\end{figure}

For glueball decay properties, no theoretical predictions are available, including from LQCD. The idea that glueball decays and charmonium decays exhibit qualitative similarities, since both proceed predominantly via gluons, was first proposed in Ref.~\cite{Chao:1995hd}. In this picture, the decays of the $0^{++}$ and $2^{++}$ glueballs should be similar to those of $\chi_{c0}$ and $\chi_{c2}$, respectively~\cite{Chao:1995hd}. By analogy, the decays of a $0^{-+}$ glueball are expected to resemble those of the $\eta_{c}$~\cite{Huang:1995td,Huang:2025pyv,Morningstar:2024vjk}, as illustrated in Fig.~\ref{fig:XDecay}. As a conventional $c\bar c$ charmonium state, the $\eta_c$ can also decay electromagnetically via photons, although its hadronic decays via gluons dominate. Furthermore, since the $0^{-+}$ glueball and the $\eta_c$ may have different masses, the resulting phase-space differences make a direct quantitative comparison of their decay properties difficult.

The $KK\pi$ and $\pi\pi\eta$ modes are two typical decay channels of the $\eta_{c}$. In addition, the decay $\eta_{c} \to a_{0}(980)\pi$ has been observed, where the $a_{0}(980)$ dominantly decays into $\pi\eta$~\cite{pdg2024}. Therefore, it is essential to search for corresponding decay modes of the $X(2370)$, such as $KK\pi$ and $\pi\pi\eta$, including the $a_{0}(980)\pi$ intermediate state, in order to investigate whether it can be interpreted as a pseudoscalar glueball. In this work, the $J/\psi\rightarrow\gamma K^{0}_{S}K^{0}_{S}\pi^{0}$ and $J/\psi\rightarrow\gamma \pi^{0}\pi^{0}\eta$ processes are studied. These channels benefit from very low background contributions because the processes $J/\psi\rightarrow\pi^{0} K^{0}_{S}K^{0}_{S}\pi^{0}$ and $J/\psi\rightarrow\pi^{0} \pi^{0}\pi^{0}\eta$ are forbidden by $C$-parity conservation and the exchange symmetry of identical particles ($K^{0}_{S}K^{0}_{S}$ and $\pi^{0}\pi^{0}$). This study is based on $(10087\pm44)\times10^{6}$ $J/\psi$ events~\cite{number} collected with the BESIII detector~\cite{detector}.

A detailed description of the design and performance of the BESIII detector can be found in Ref.~\cite{detector}. Monte Carlo (MC) simulated data samples produced with a {\sc geant4}-based~\cite{geant4} package, which includes the geometric description of the BESIII detector~\cite{detector2022} and the detector response, are used to determine detection efficiencies and estimate backgrounds. The simulation models the beam energy spread and initial-state radiation in $e^+e^-$ annihilations with the generator {\sc kkmc}~\cite{KKMC,KKMC2}. The inclusive MC sample includes both the production of the $J/\psi$ resonance and the continuum processes incorporated in {\sc kkmc}~\cite{KKMC,KKMC2}. All particle decays are modeled with {\sc evtgen}~\cite{EVTGEN,EVTGEN2} using branching fractions either taken from the Particle Data Group (PDG)~\cite{pdg2024}, when available, or otherwise estimated with {\sc lundcharm}~\cite{Lund-Charm,Lund-Charm2}. Final-state radiation from charged particles is incorporated using the {\sc photos} package~\cite{PHOTOS}. MC simulation samples of the processes $J/\psi \rightarrow \gamma K^{0}_{S}K^{0}_{S}\pi^{0}$ and $J/\psi \rightarrow \gamma \pi^{0}\pi^{0}\eta$, including the subsequent decays $K^{0}_{S} \rightarrow \pi^{+}\pi^{-}$, $\pi^{0} \to \gamma\gamma$, and $\eta \to \gamma\gamma$, are generated uniformly in phase-space (PHSP).

Charged tracks detected in the multi-layer drift chamber (MDC) are required to be within a polar-angle ($\theta$) range of $|\rm{cos\theta}|<0.93$, where $\theta$ is defined with respect to the $z$-axis, the symmetry axis of the MDC. No requirement is imposed on the distance of closest approach to the interaction point for charged tracks since all charged tracks originate from $K^{0}_{S}$ decays. All charged tracks are assumed to be pions. To reconstruct $K_{S}^{0}$ candidates, the tracks of each $\pi^{+}\pi^{-}$ pair are fitted to a secondary vertex. To suppress background events, all $K_{S}^{0}$ candidates are required to satisfy $|M_{\pi^{+}\pi^{-}}-m_{K_{S}^{0}}|<11~\MeV/c^{2}$, where $m_{K_{S}^{0}}$ is the nominal mass of the $K_{S}^{0}$ from the PDG~\cite{pdg2024}. To further suppress background, the decay length of the $K_{S}^{0}$ candidate is required to be greater than twice the vertex resolution away from the interaction point. With these selections, the miscombination rate between pions from different $K_{S}^{0}$ candidates is 0.2\%. The reconstructed $K_{S}^{0}$ candidates are used as input for the subsequent kinematic fit.

Photon candidates are identified using isolated showers in the electromagnetic calorimeter (EMC). The deposited energy of each shower must be greater than 25~MeV in the barrel region ($|\cos \theta|< 0.80$) and greater than 50~MeV in the end cap region ($0.86 <|\cos \theta|< 0.92$). To exclude showers that originate from charged tracks, the opening angle between the shower position and charged tracks extrapolated to the EMC must be greater than 10 degrees. The EMC time of a shower is required to be within 700 ns after the $e^{+}e^{-}$ annihilation (the time determined by charged tracks), or, in the case of no charged tracks, within 500 ns before and after the time of the highest-energy shower.

\begin{figure}[tb]
\centering
\vspace{-5mm}
\subfloat{
\hspace{-2mm}
  \includegraphics[width=0.5\textwidth]{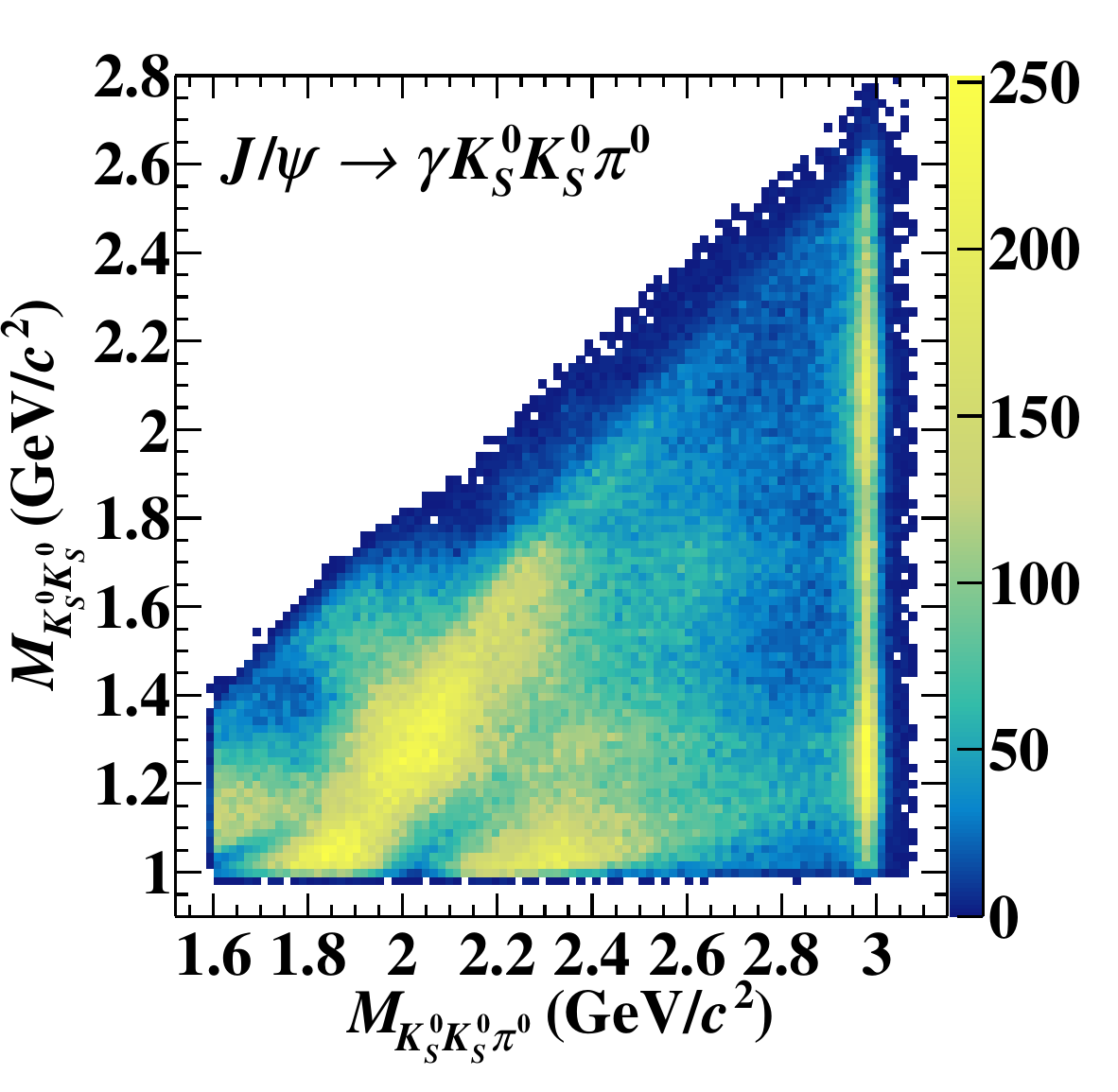}
  \label{fig:plot_mkspi0_mkskspi0}
\put(-100,86){(a)}
}\subfloat{
\hspace{-2.5mm}
  \includegraphics[width=0.5\textwidth]{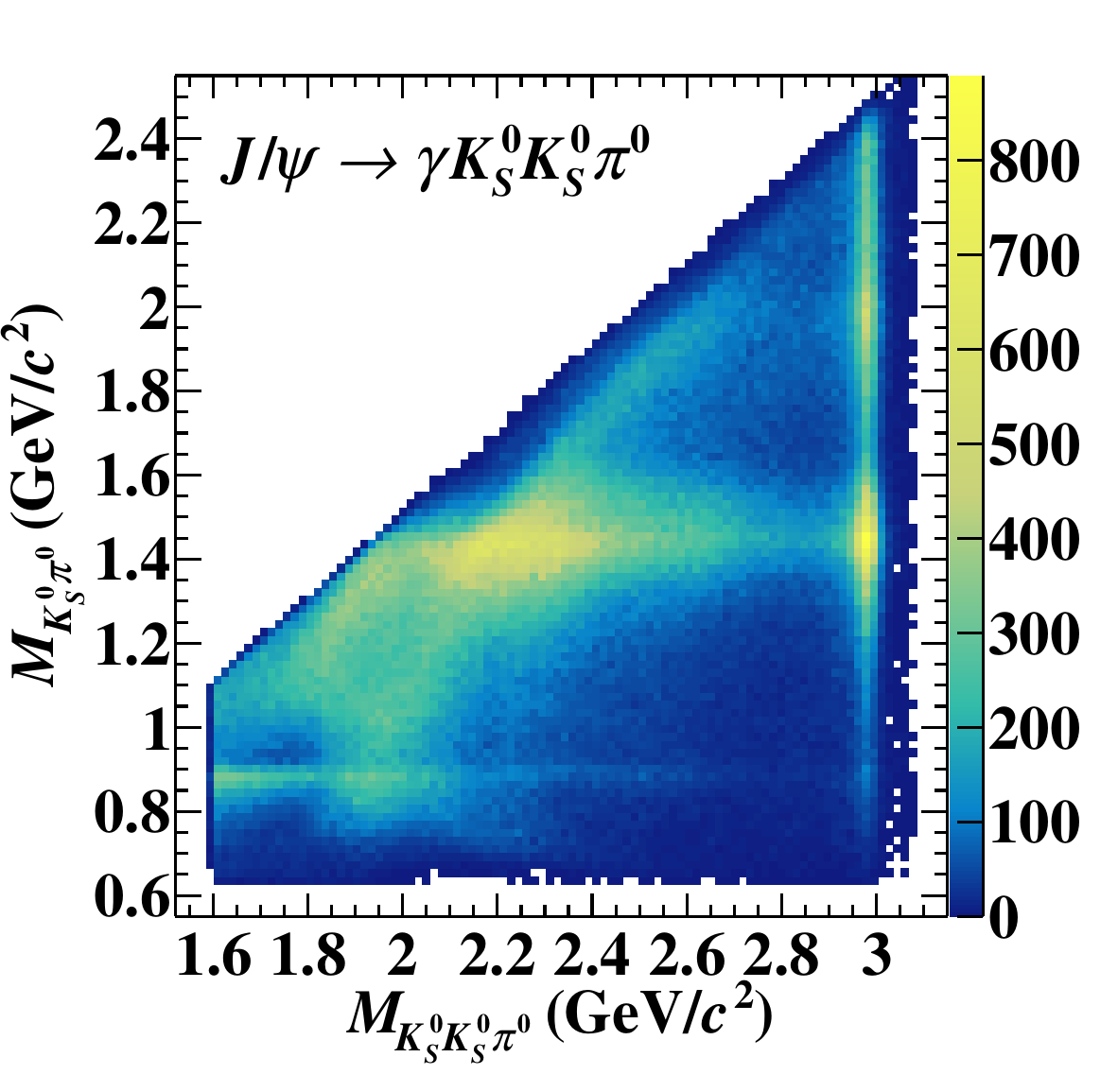}
  \label{fig:plot_mksks_mkskspi0}
\put(-100,86){(b)}
}\\
\vspace{-4.5mm}
\subfloat{
\hspace{-2mm}
  \includegraphics[width=0.5\textwidth]{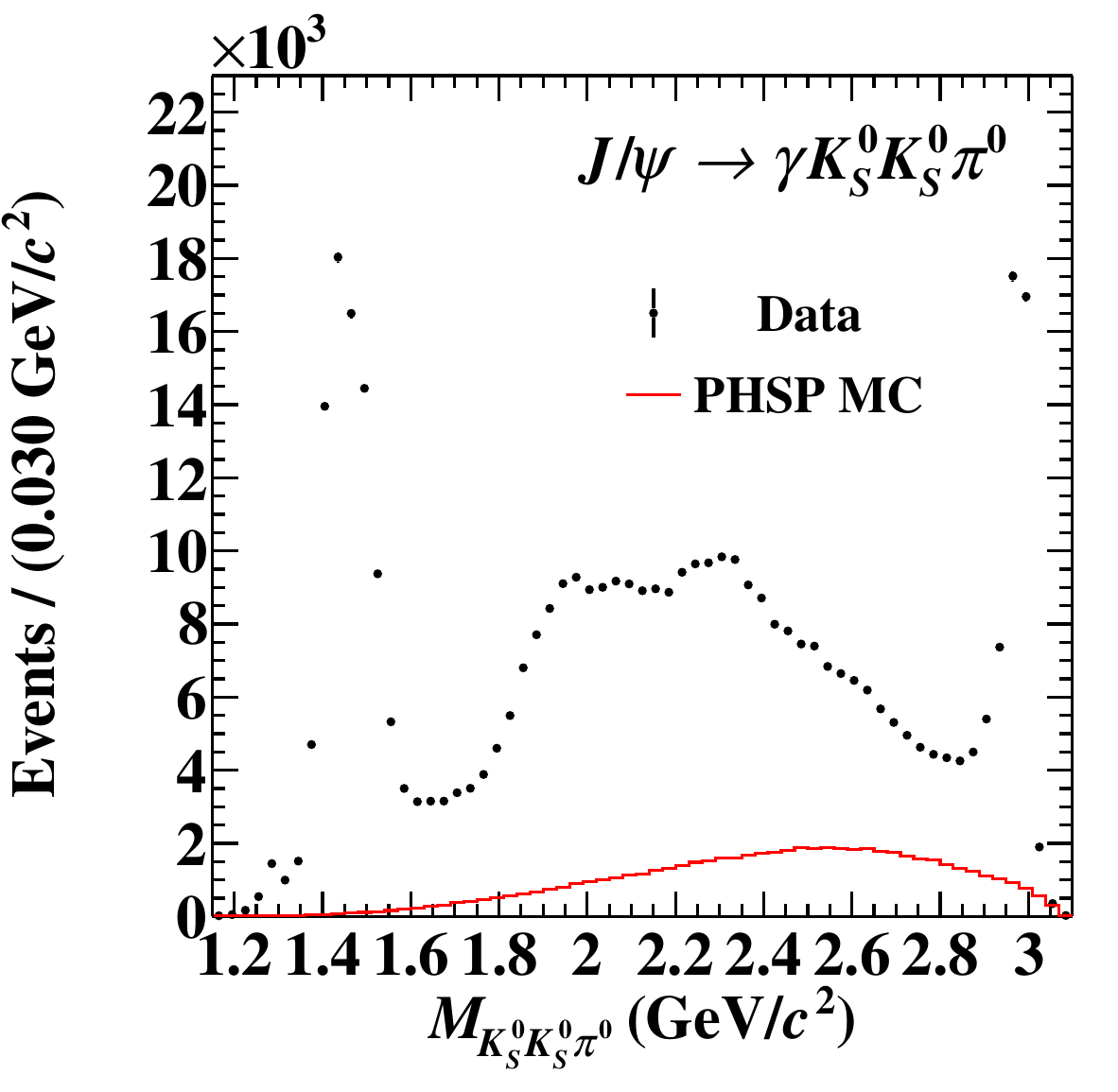}
  \label{fig:plot_mkskspi0}
\put(-100,95){(c)}
}
\subfloat{
\hspace{-4mm}
  \includegraphics[width=0.5\textwidth]{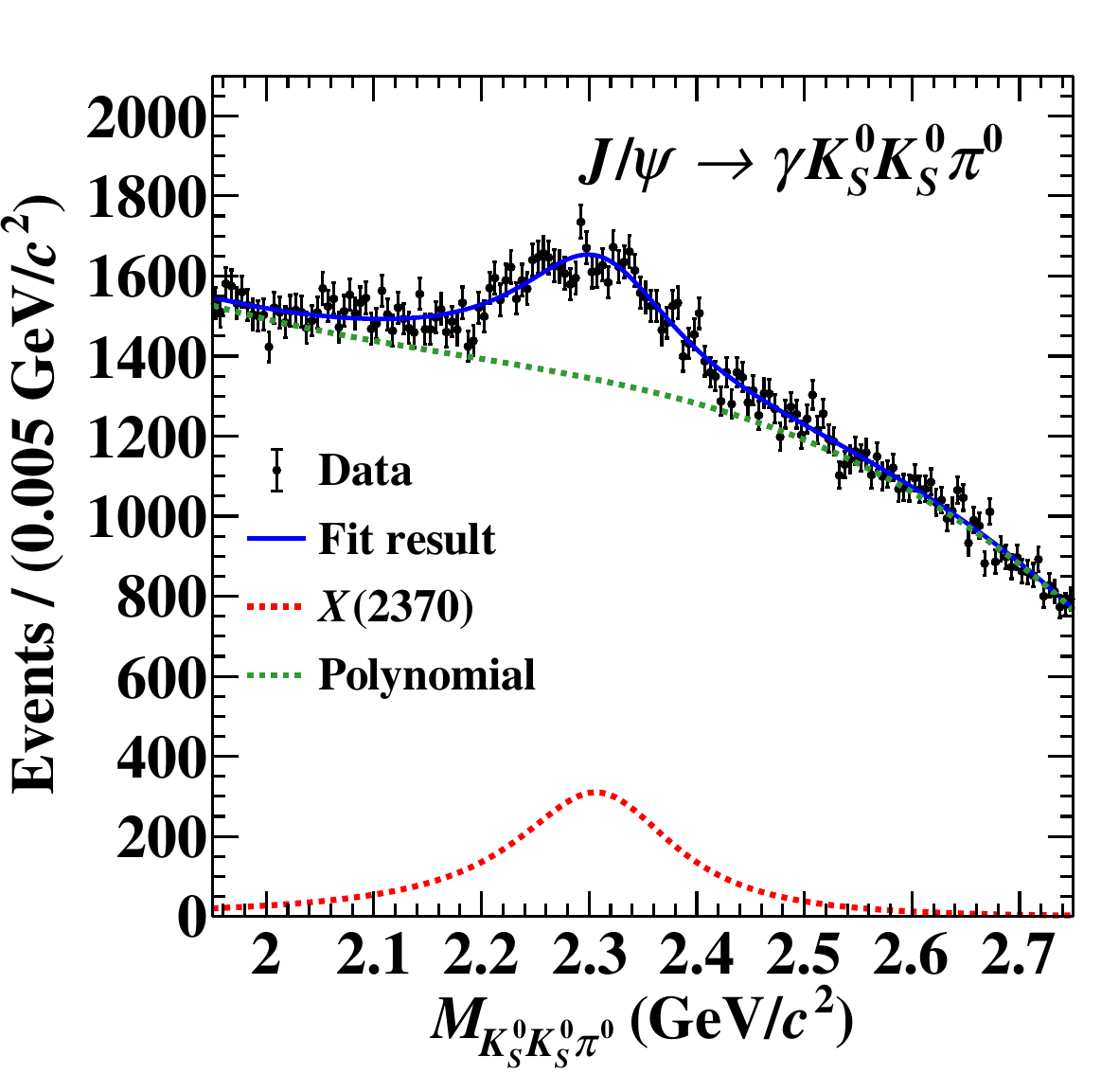}
  \label{fig:fit_mkskspi0} 
\put(-100,95){(d)}
}\\
\vspace{-3.0mm}
\caption{
    Two-dimensional distributions of (a) $M_{K^{0}_{S}K^{0}_{S}}$ versus $M_{K^{0}_{S}K^{0}_{S}\pi^{0}}$ and (b) $M_{K^{0}_{S}\pi^{0}}$ versus $M_{K^{0}_{S}K^{0}_{S}\pi^{0}}$. 
(c) The $K^{0}_{S}K^{0}_{S}\pi^{0}$ invariant mass spectrum. The solid red histogram shows PHSP MC with arbitrary normalization.
(d) The fit results within the range [1.95, 2.75]~$\GeV/c^2$. The dashed red line is the $X(2370)$ signal component, the dashed green line represents the remaining processes described by a third-order polynomial function, and the blue line is the total fit result. 
}
  \label{fig:spectra_and_fit1}
\end{figure}

For the $J/\psi\rightarrow\gamma K_{S}^{0}K_{S}^{0}\pi^{0}$ process, each event candidate is required to have at least two $K_{S}^{0}$ candidates, at least three photons, and no additional charged tracks from the interaction point. A four-constraint (4C) kinematic fit under the $J/\psi\rightarrow\gamma\gamma\gamma K_{S}^{0}K_{S}^{0}$ hypothesis is performed by enforcing energy-momentum conservation. If there are multiple $\gamma\gamma\gamma K_{S}^{0}K_{S}^{0}$ combinations, the one with the smallest $\chi^{2}_{\mathrm{4C}}$ is chosen. The resulting $\chi^{2}_{\mathrm{4C}}$ is required to be less than 40. To reconstruct the $\pi^{0}$ candidate, a five-constraint (5C) kinematic fit is performed to further constrain the invariant mass of the two photons to $m_{\pi^{0}}$, where $m_{\pi^{0}}$ is the nominal mass of the $\pi^{0}$~\cite{pdg2024}. Among the three $\gamma\gamma$ combinations, the one with the smallest $\chi^{2}_{\mathrm{5C}}$ is chosen as the $\pi^{0}$ candidate. The two photons from the $\pi^{0}$ candidate are required to satisfy $|M_{\gamma\gamma} - m_{\pi^{0}}| < 22~\MeV/c^{2}$, with each photon having an energy greater than $100~\MeV$. To suppress miscombinations of the radiative photon ($\gamma_{\text{rad}}$) and the photons from the $\pi^{0}$ candidate ($\gamma_{\pi^{0}}$), events with $|M_{\gamma_{\text{rad}}\gamma_{\pi^{0}}} - m_{\pi^{0}}| < 22~\MeV/c^{2}$ are rejected. These photon-related requirements reduce the miscombination rate between the $\gamma_{\text{rad}}$ and $\gamma_{\pi^{0}}$ to $0.1\%$. Events with $|M_{\gamma_{\text{rad}}\gamma_{\pi^{0}}} - m_{\eta}| < 25~\MeV/c^{2}$ or $|M_{\gamma_{\text{rad}}\pi^{0}} - m_{\omega}| < 40~\MeV/c^{2}$ are rejected to suppress background containing an $\eta$ or $\omega$, where $m_{\eta}$ and $m_{\omega}$ are the nominal masses of the $\eta$ and $\omega$~\cite{pdg2024}, respectively. After applying the above selection criteria, a clear signal of the $X(2370)$ around $2.3~\GeV/c^{2}$, along with the $\eta_{c}$ peak around $2.9~\GeV/c^{2}$, is observed in the $K^{0}_{S}K^{0}_{S}\pi^{0}$ invariant-mass spectrum using the momenta obtained from the 5C kinematic fit, as shown in Fig.~\ref{fig:plot_mkspi0_mkskspi0}, Fig.~\ref{fig:plot_mksks_mkskspi0}, and Fig.~\ref{fig:plot_mkskspi0}. These two-dimensional distributions indicate several possible intermediate processes, for example, $a_{0}(980)^{0}\pi^{0}$ and $K^{*}_{0}(1430)K^{0}_{S}\to K^{0}_{S}K^{0}_{S}\pi^{0}$.

An inclusive MC sample of 10 billion $J/\psi$ decays is used to study potential background contributions. After applying the above selection criteria to the inclusive MC sample, no peaking background is found in the $K^{0}_{S}K^{0}_{S}\pi^{0}$ invariant-mass spectrum. The contributions of non-$K^{0}_{S}$ and non-$\pi^{0}$ background processes are found to be negligible in the data by fitting the corresponding $M_{\pi^{+}\pi^{-}}$ and $M_{\gamma\gamma}$ distributions.

An unbinned maximum-likelihood fit is performed on the $K^{0}_{S}K^{0}_{S}\pi^{0}$ invariant-mass spectrum between $1.95$ and $2.75~\GeV/c^{2}$. The $X(2370)$ signal is described by an efficiency-corrected and PHSP-weighted Breit-Wigner (BW) function convoluted with a detector resolution function. A third-order polynomial function is used to describe the remaining contributions. From the fit shown in Fig.~\ref{fig:fit_mkskspi0}, the mass and width of the $X(2370)$ are measured to be $2311\pm4(\mathrm{stat})~\MeV/c^{2}$ and $177\pm20(\mathrm{stat})~\MeV$, respectively. The significance of the $X(2370)$ is greater than $14\sigma$, as determined from the changes in the log-likelihood value and degrees of freedom between fits with and without the $X(2370)$ signal hypothesis, and includes systematic variations.

\begin{figure}[tb]
\centering
\vspace{-5mm}
\subfloat{
\hspace{-2mm}
  \includegraphics[width=0.5\textwidth]{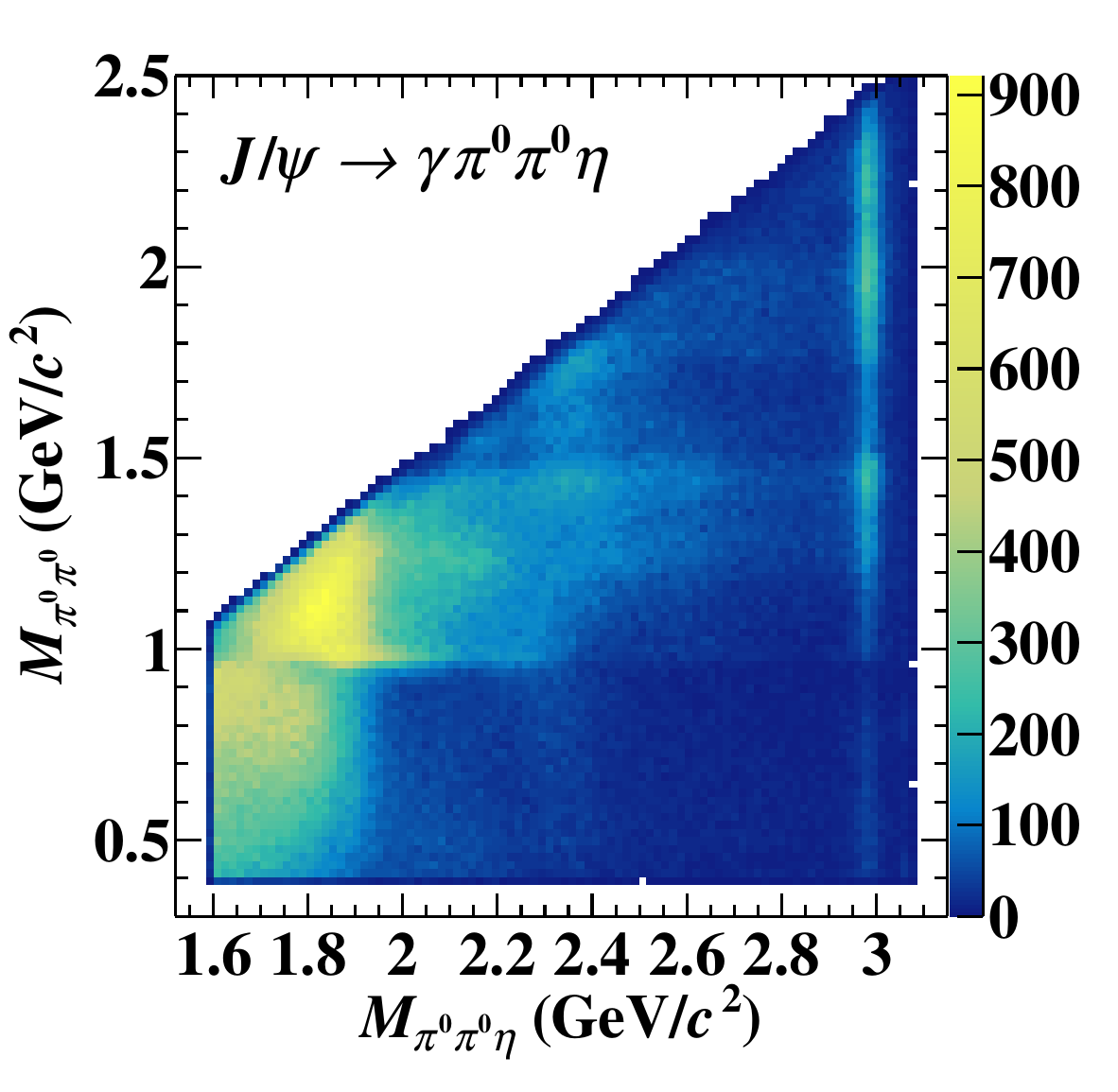}
  \label{fig:plot_mpi0eta_mpi0pi0eta}
\put(-100,86){(a)}
}
\subfloat{
\hspace{-2.5mm}
  \includegraphics[width=0.5\textwidth]{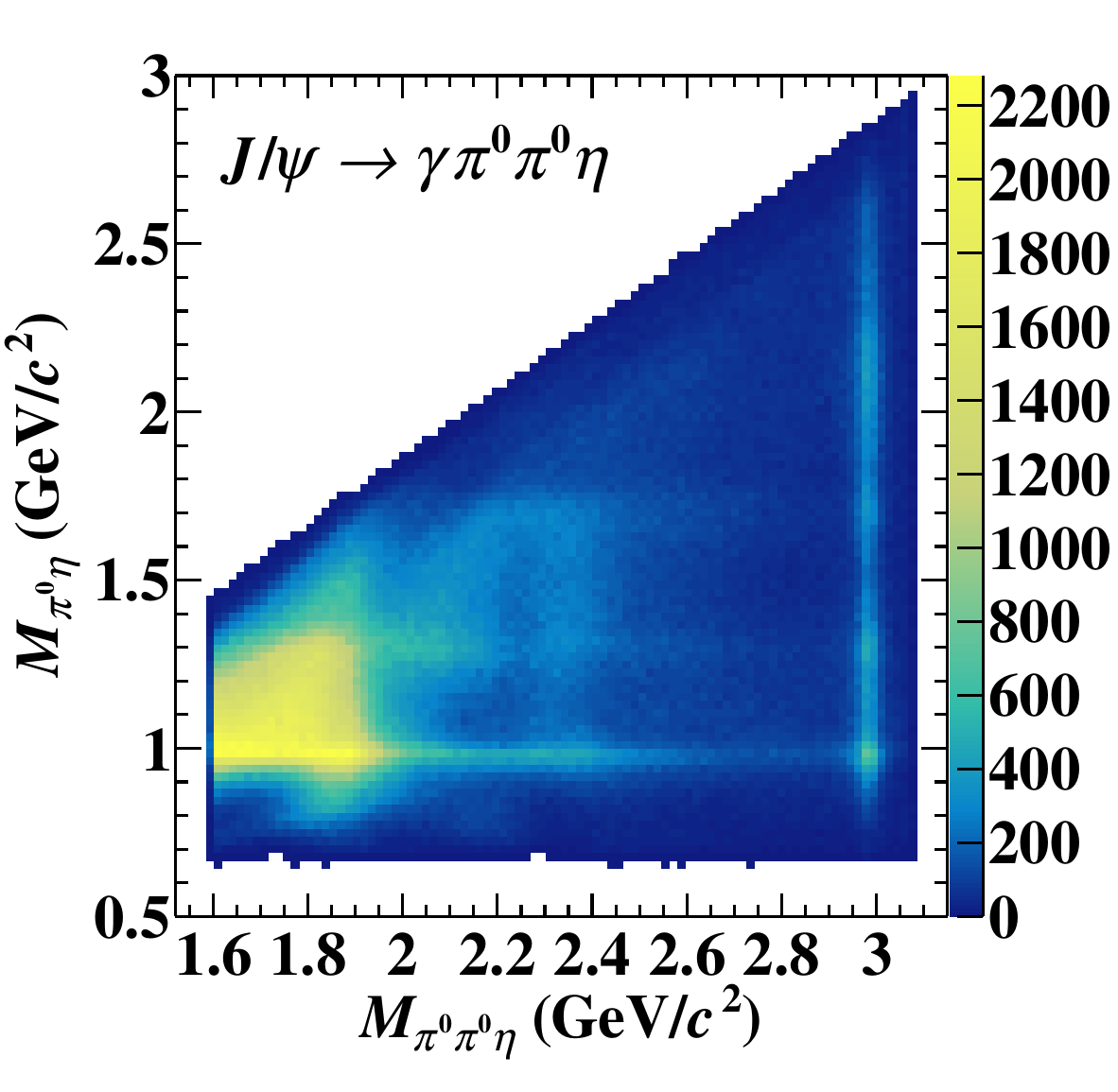}
  \label{fig:plot_mpi0pi0_mpi0pi0eta}
\put(-100,86){(b)}
}\\
\vspace{-4.5mm}
\subfloat{
\hspace{-2mm}
  \includegraphics[width=0.5\textwidth]{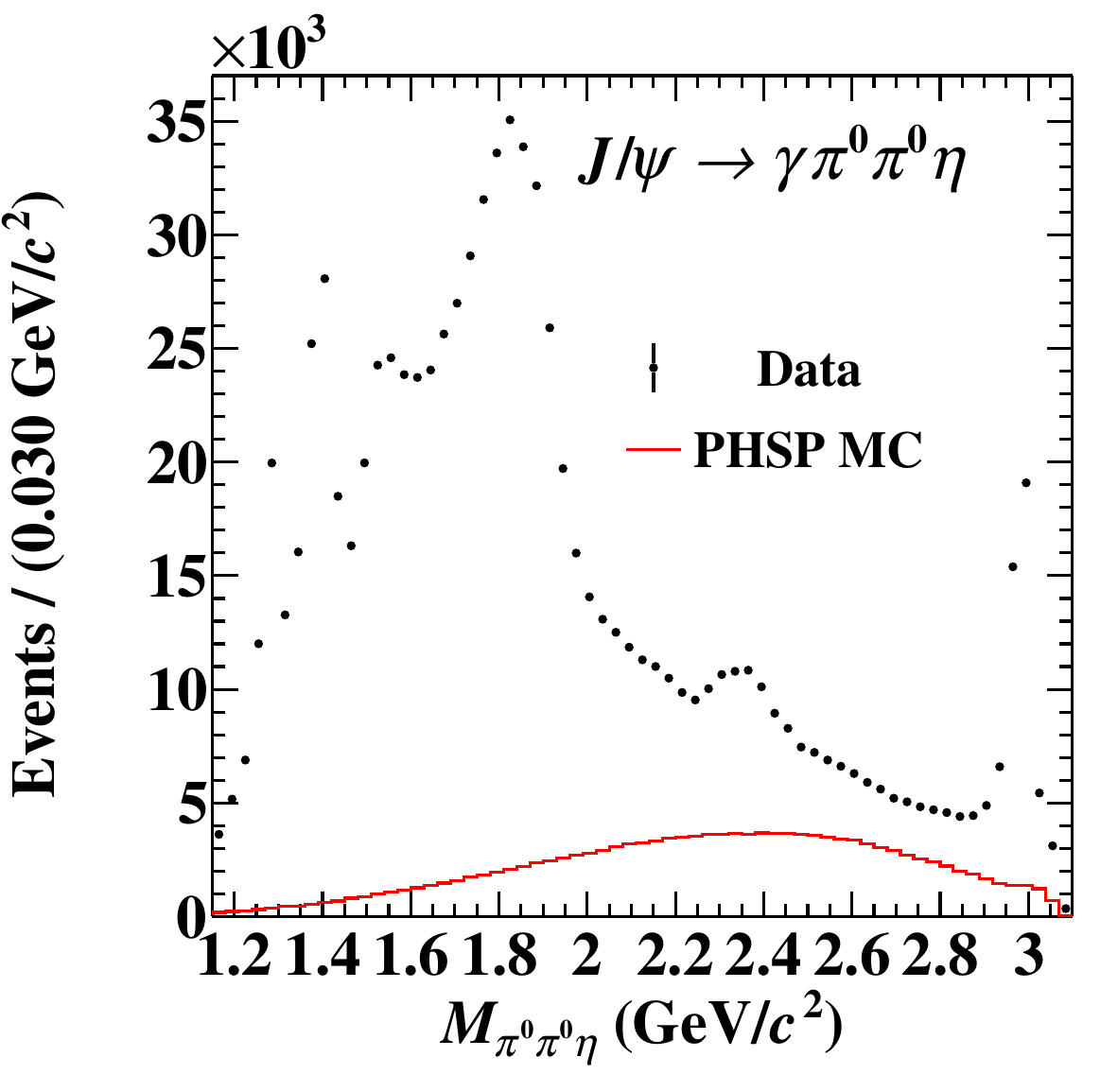}
  \label{fig:plot_mpi0pi0eta} 
\put(-98,98){(c)}
}
\subfloat{
\hspace{-4mm}
  \includegraphics[width=0.5\textwidth]{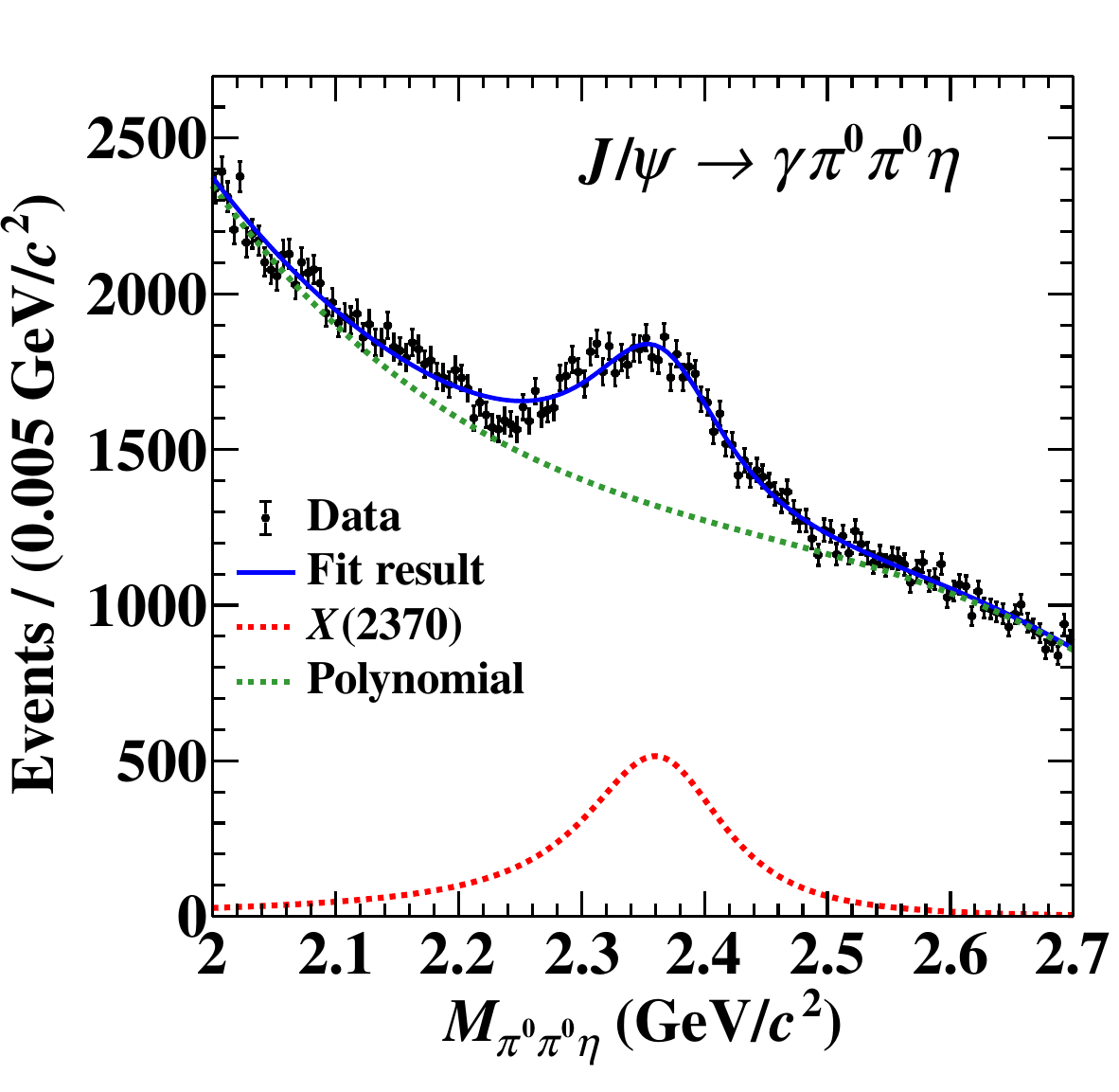}
  \label{fig:fit_mpi0pi0eta}
\put(-98,98){(d)}
}
\vspace{-3.0mm}

\caption{
The two-dimensional distributions of (a) $M_{\pi^{0}\pi^{0}}$ versus $M_{\pi^{0}\pi^{0}\eta}$ and (b) $M_{\pi^{0}\eta}$ versus $M_{\pi^{0}\pi^{0}\eta}$. 
(c) The $\pi^{0}\pi^{0}\eta$ invariant-mass spectrum. The solid red histogram corresponds to the PHSP MC sample with arbitrary normalization.
(d) The fit results in the range [2.00, 2.70]~$\GeV/c^2$. The dashed red line represents the $X(2370)$ signal component, the dashed green line represents the remaining processes described by a third-order polynomial function, and the blue line indicates the total fit result.
}
  \label{fig:spectra_and_fit2}
\end{figure}

For the $J/\psi\rightarrow\gamma \pi^{0}\pi^{0}\eta$ process, 
each event candidate is required to have no charged tracks and at least seven photons.
To identify photon pairs originating from $\pi^{0}$ decays,
a one-constraint (1C) kinematic fit is performed by
constraining the invariant mass of each photon pair to $m_{\pi^{0}}$.
At least two pairs with $\chi^{2}_{\mathrm{1C}}<10$ are required and selected as $\pi^{0}$ candidates, 
which are then used as inputs to the subsequent kinematic fit.
A six-constraint (6C) kinematic fit under the $J/\psi\rightarrow \gamma\gamma\gamma\pi^{0}\pi^{0}$ hypothesis is performed 
by enforcing energy-momentum conservation and constraining 
the invariant masses of the two $\pi^{0}$ candidates to $m_{\pi^{0}}$.
If there are multiple $\gamma\gamma\gamma\pi^{0}\pi^{0}$ combinations, 
the one with the smallest $\chi^{2}_{\mathrm{6C}}$ is chosen. 
The resulting $\chi^{2}_{\mathrm{6C}}$ is required to be less than 40. 
To reconstruct the $\eta$ candidate, a seven-constraint (7C) kinematic fit is performed to further
constrain the invariant mass of the two photons to $m_{\eta}$.
Among the three $\gamma\gamma$ combinations, the one with the smallest $\chi^{2}_{\mathrm{7C}}$ is chosen.
To suppress multi-photon backgrounds,
the $\chi^{2}_{4C}$ from the kinematic fit under the $J/\psi\to 7 \gamma$ hypothesis is required to be less than that from the kinematic fits under the $J/\psi\to 8 \gamma$ and $J/\psi\to 9 \gamma$ hypotheses. 
The two photons from the $\eta$ candidate are required to satisfy $|M_{\gamma\gamma} - m_{\eta}| < 27~\MeV/c^{2}$. 
To suppress miscombination between the radiative photon and other selected photons,
events satisfying any of the following requirements are rejected:
$|M_{\gamma_{\text{rad}}\gamma_{\pi^{0}}} - m_{\pi^{0}}|<20~\MeV/c^{2}$,
$|M_{\gamma_{\text{rad}}\gamma_{\eta}} - m_{\pi^{0}}|<20~\MeV/c^{2}$,
$|M_{\gamma_{\text{rad}}\gamma_{\pi^{0}}}-m_{\eta}|<30~\MeV/c^{2}$, and $|M_{\gamma_{\text{rad}}\gamma_{\eta}}-m_{\eta}|<50~\MeV/c^{2}$,
where $\gamma_{\eta}$ denotes the photons from the $\eta$ candidate.
With these requirements, the miscombination rate among the seven photons is $1.4\%$. 
To suppress background containing an $\omega$, events with $|M_{\gamma_{\text{rad}}\pi^{0}} - m_{\omega}| < 40~\MeV/c^{2}$ are rejected.
After applying the above selection criteria,
clear $X(2370)$ and $\eta_{c}$ peaks are observed in the $\pi^{0}\pi^{0}\eta$ invariant mass spectrum using the momenta obtained from the 7C kinematic fit, 
as shown in Fig.~\ref{fig:plot_mpi0eta_mpi0pi0eta}, Fig.~\ref{fig:plot_mpi0pi0_mpi0pi0eta}, and Fig.~\ref{fig:plot_mpi0pi0eta}.
These two-dimensional distributions also indicate several possible intermediate processes, such as $f_{0}(1500)\eta$ and $a_{0}(980)^{0}\pi^{0}\to\pi^{0}\pi^{0}\eta$.

After applying the above selection criteria, 
no peaking background is found in the $\pi^{0}\pi^{0}\eta$ invariant mass spectrum of the inclusive MC sample. 
The background contribution from non-$\eta$ processes is estimated by fitting the corresponding $M_{\gamma\gamma}$ distribution in data, yielding a non-$\eta$ background fraction of 3.5\%. 
A validation using $\eta$ mass sideband events indicates that the non-$\eta$ background processes produce no peaking structure. 
Therefore, it is described by a polynomial function in the subsequent fit.

A strategy analogous to that described for the $J/\psi\to\gamma K^{0}_{S}K^{0}_{S}\pi^{0}$ process is employed in the fit to the $\pi^{0}\pi^{0}\eta$ invariant mass spectrum between $2.0$ and $2.7~\GeV/c^{2}$, as shown in Fig.~\ref{fig:fit_mpi0pi0eta}. 
The mass and width of the $X(2370)$ are measured to be $2366\pm2(\mathrm{stat})~\MeV/c^{2}$ and $127\pm 9(\mathrm{stat})~\MeV$, respectively.
The significance of the $X(2370)$ is greater than $20\sigma$, including systematic variations.

\begin{figure}[tb]
\centering
\vspace{-5mm}

\subfloat{
\hspace{-2mm}
  \includegraphics[width=0.5\textwidth]{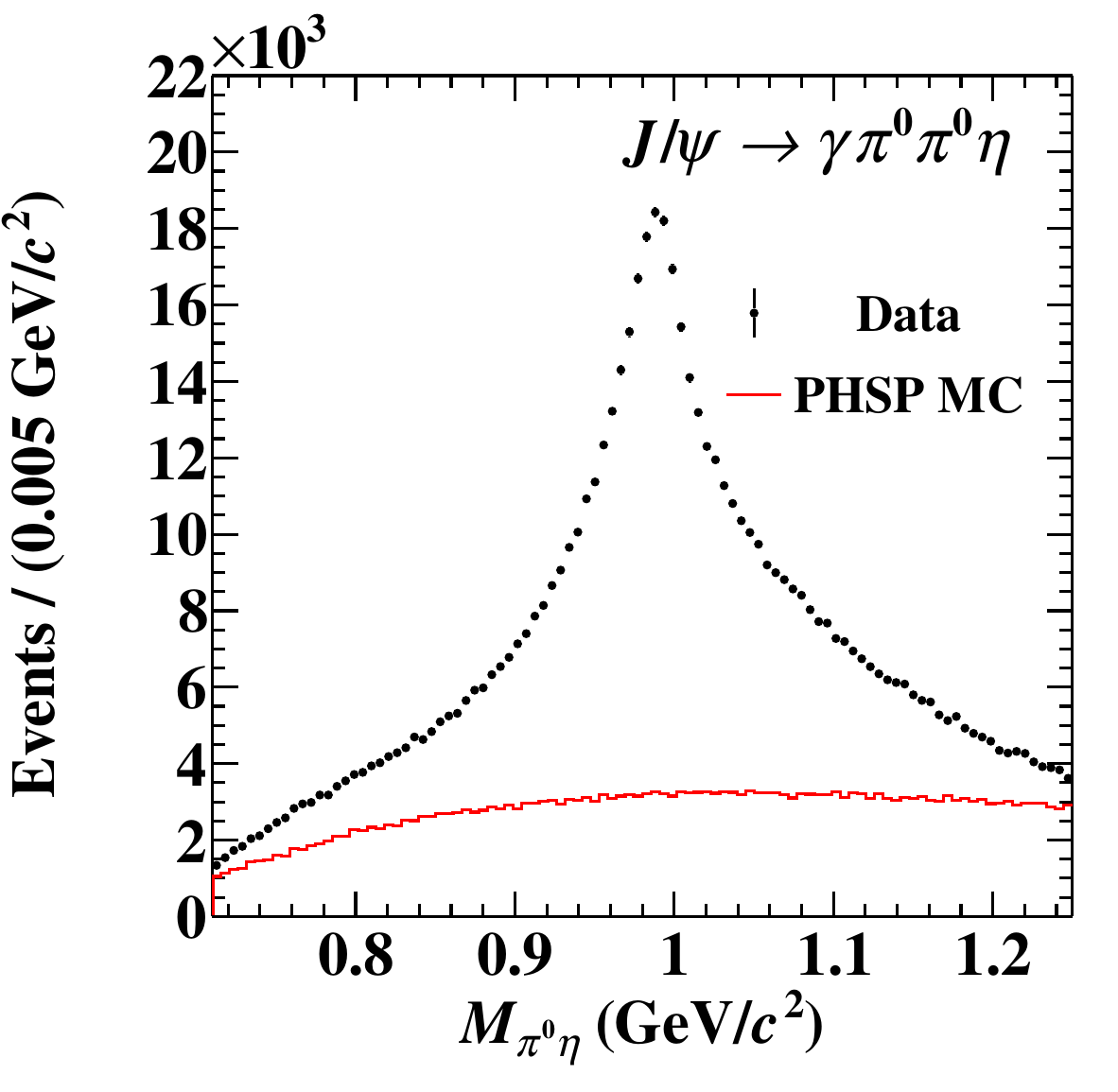}
  \label{fig:plot_mpi0eta_a0} 
\put(-100,95){(a)}
} 
\subfloat{
\hspace{-3.5mm}
  \includegraphics[width=0.5\textwidth]{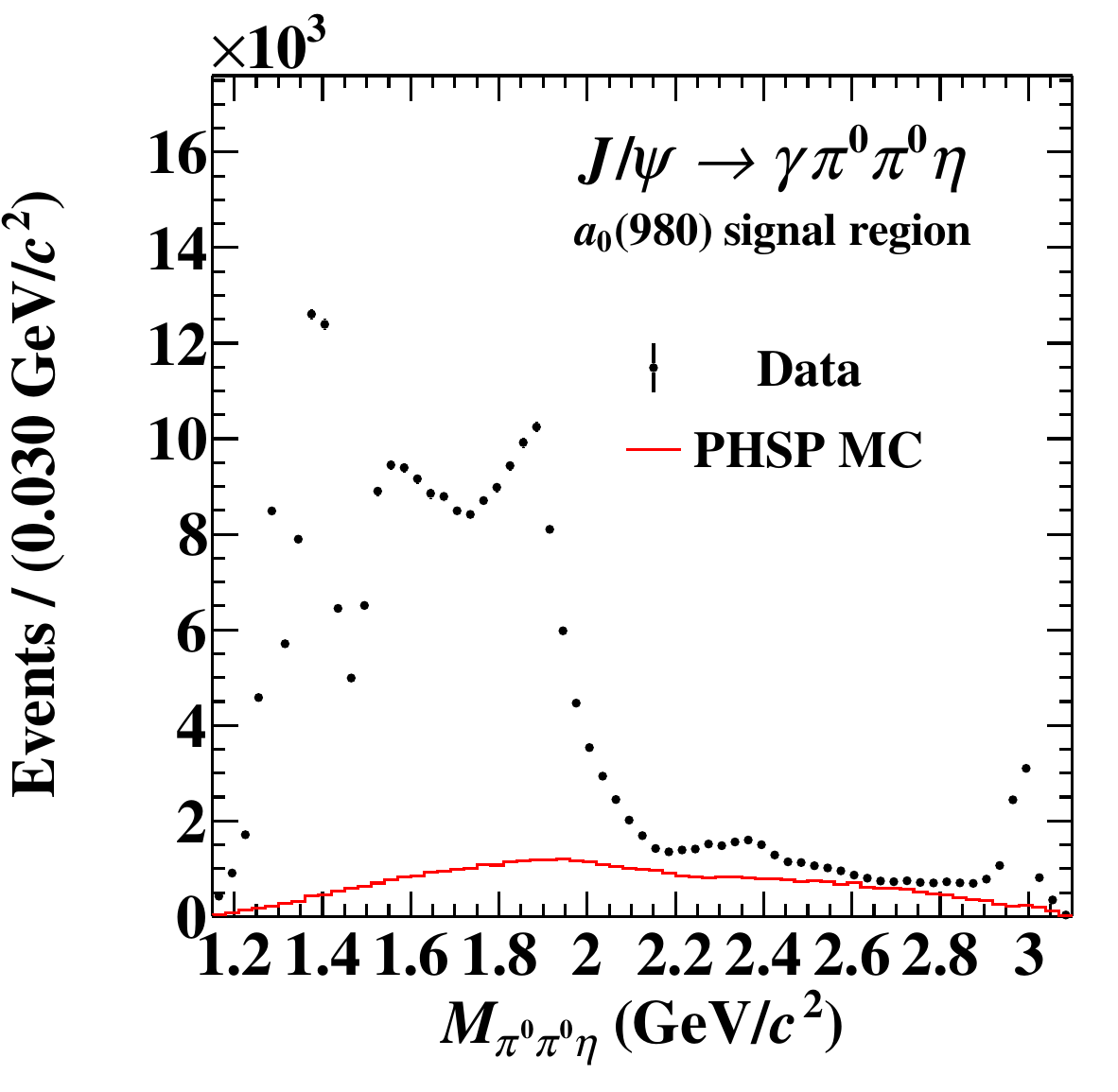}
  \label{fig:plot_mpi0pi0eta_a0} 
\put(-100,95){(b)}
}
\vspace{-4.5mm}
\subfloat{
\hspace{-2mm}
  \includegraphics[width=0.5\textwidth]{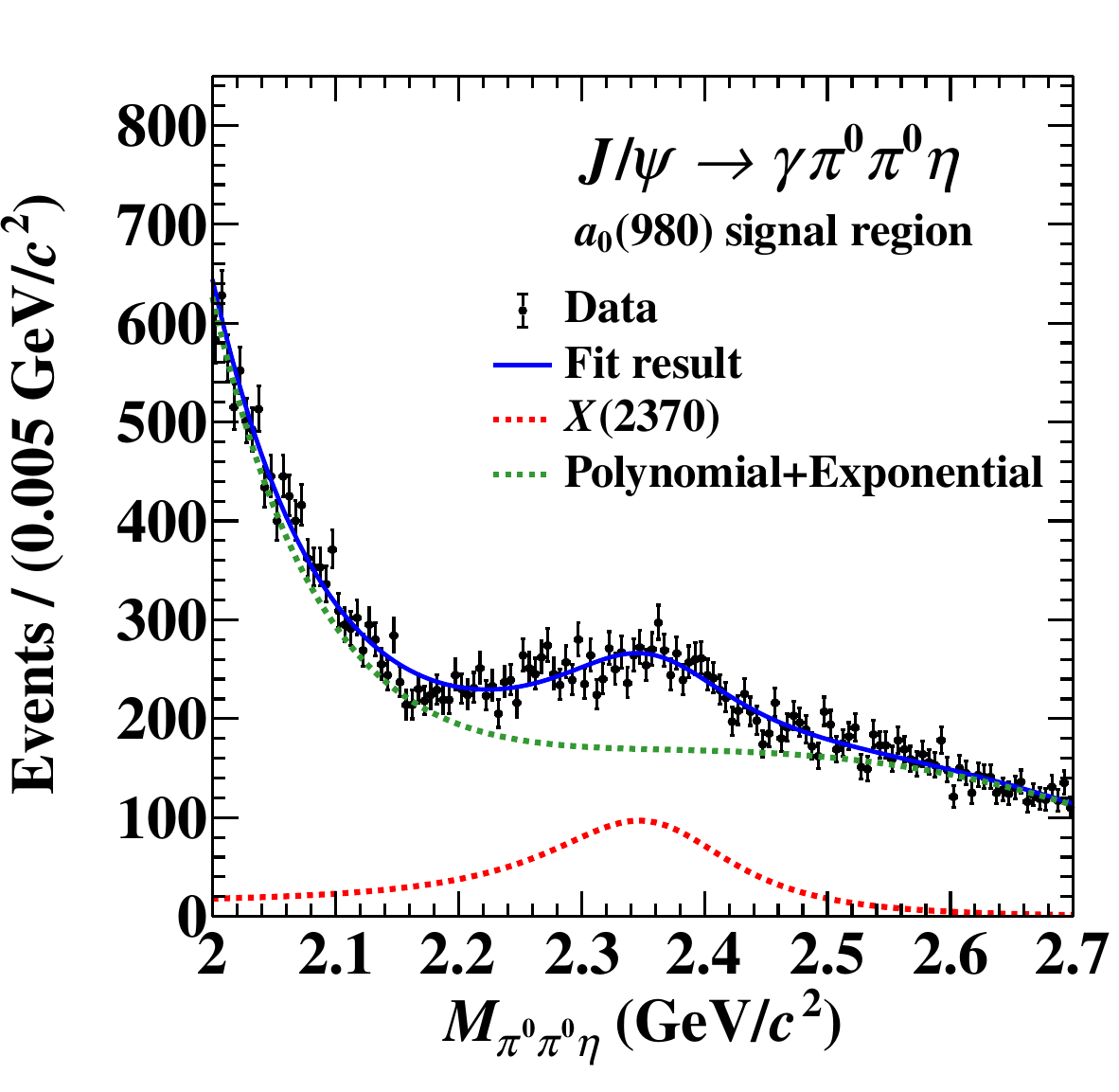}
  \label{fig:fit_mpi0pi0eta_a0_signal}
\put(-100,95){(c)}
}
\subfloat{
\hspace{-3.5mm}
  \includegraphics[width=0.5\textwidth]{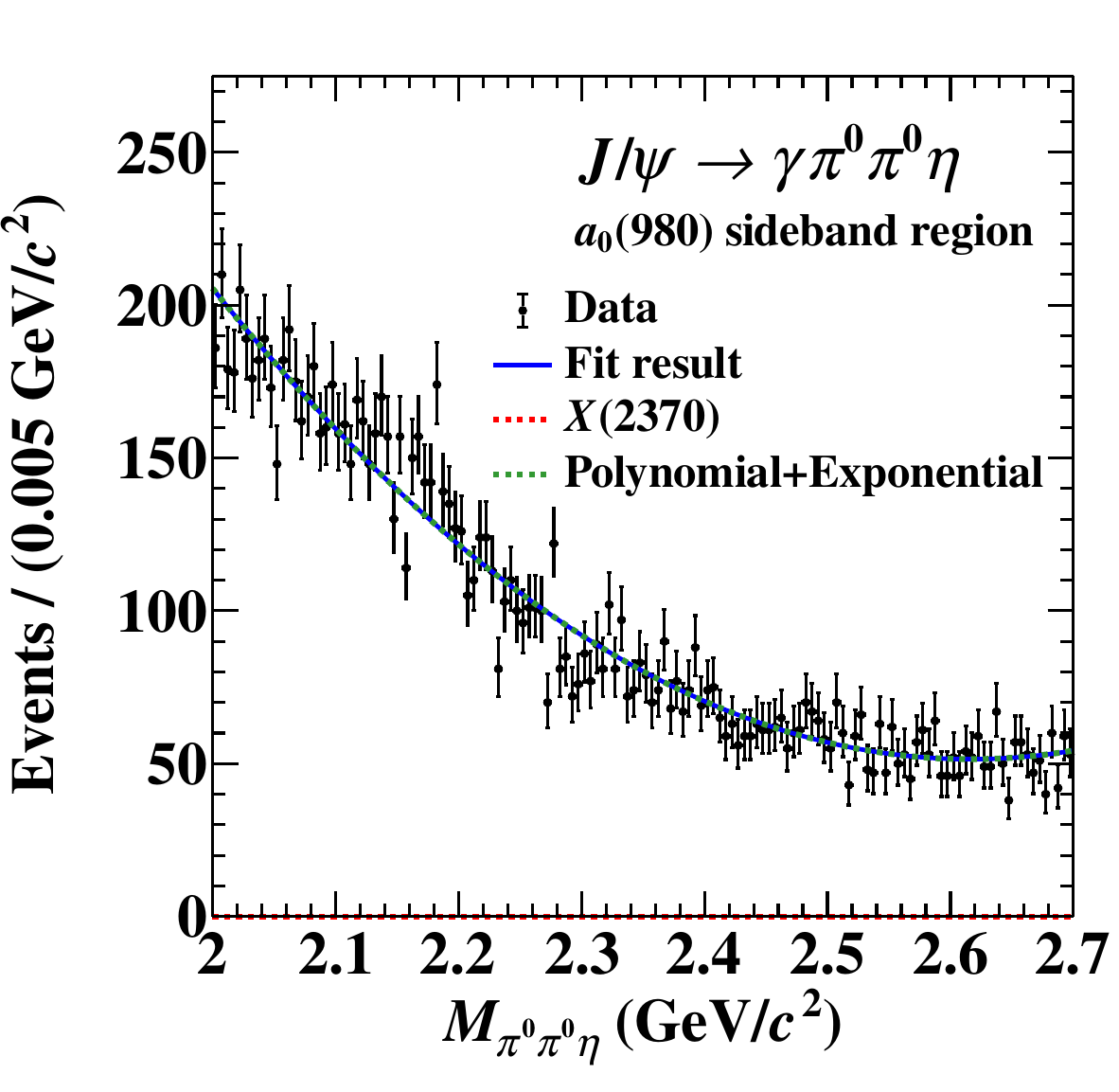}
  \label{fig:fit_mpi0pi0eta_a0_sideband}
\put(-100,95){(d)}
}
\vspace{-3.0mm}
\caption{
(a) The invariant mass spectrum of the $\eta$ and the lower-energy $\pi^{0}$. 
(b) The $\pi^{0}\pi^{0}\eta$ invariant mass spectrum for selected events in the $a_{0}(980)^{0}$ signal region. 
The PHSP MC simulations are shown with arbitrary normalization. 
(c) and (d) show the fit results in the range [2.00, 2.70]~$\GeV/c^2$ for the $a_{0}(980)^{0}$ signal region and the $a_{0}(980)^{0}$ sideband region, respectively. 
The dashed green line represents the contribution of the remaining processes, described by the sum of a second-order polynomial and an exponential function. 
}
  \label{fig:spectra_and_fit5}
\end{figure}

Using the selection criteria for the $J/\psi\to\gamma \pi^{0}\pi^{0}\eta$ process described above, an $a_{0}(980)^{0}$ candidate is reconstructed from the $\eta$ and the lower-energy $\pi^{0}$. To suppress contributions from the decay $X(2370)\to f_{0}(1500)\eta$, events with $|M_{\pi^{0}\pi^{0}} - 1.5~\GeV/c^2 |< 0.15~\GeV/c^2$ are rejected. The resulting mass spectrum of the $\eta$ and the lower-energy $\pi^{0}$ is shown in Fig.~\ref{fig:plot_mpi0eta_a0}. An additional requirement of $|M_{\pi^{0}\eta}-0.98~\GeV/c^2|<0.05~\GeV/c^2$ is applied to select the $a_{0}(980)^{0}$ signal region. After applying these requirements, the $\pi^{0}\pi^{0}\eta$ mass spectrum, shown in Fig.~\ref{fig:plot_mpi0pi0eta_a0}, exhibits clear $X(2370)$ and $\eta_{c}$ peaks. The $\pi^{0}\pi^{0}\eta$ mass spectrum between $2.00$ and $2.70~\GeV/c^{2}$ is fitted using a strategy analogous to that described in the previous paragraph, as shown in Fig.~\ref{fig:fit_mpi0pi0eta_a0_signal}. In particular, the PHSP parametrization in this fit includes an intermediate $a_{0}(980)^{0}$ state, which is parametrized using dispersion integrals~\cite{PhysRevD.95.032002}. The remaining processes are described by the sum of a polynomial and an exponential function. The mass and width of the $X(2370)$ are measured to be $2358\pm5(\mathrm{stat})~\MeV/c^{2}$ and $180\pm 20(\mathrm{stat})~\MeV$, respectively. The significance of the $X(2370)$ is greater than $9\sigma$, including systematic variations. In contrast, there is no evidence for an $X(2370)$ signal in the $a_{0}(980)^{0}$ mass sideband region of $0.20~\GeV/c^2<|M_{\pi^{0}\eta}-0.98~\GeV/c^2|<0.25~\GeV/c^{2}$, as shown in Fig.~\ref{fig:fit_mpi0pi0eta_a0_sideband}.

Systematic uncertainties associated with the invariant-mass spectrum fits are evaluated for the three decay modes and arise primarily from the PHSP parametrization, the background model, and the treatment of an additional resonance. 
For each source, alternative fits are performed, and the largest deviations of the fitted mass and width of the $X(2370)$ from the nominal fit are taken as the corresponding systematic uncertainties. 
The uncertainty associated with the PHSP parametrization is studied by including an additional Blatt-Weisskopf barrier factor and by considering PHSP variations among different intermediate resonances. 
For the $K^{0}_{S}K^{0}_{S}\pi^{0}$ mode, 
the alternative intermediate resonances $K_{0}^{*}(1430)$ and $a_{0}(980)^{0}$ are considered. 
For the $\pi^{0}\pi^{0}\eta$ mode, the alternative intermediate resonances $a_{0}(980)^{0}$ and $f_{0}(1500)$ are considered. 
The $a_{0}(980)^{0}$ is described using both the Flatt\'e~\cite{flatte1976,PhysRevD.95.032002} and dispersion-integral parametrizations~\cite{PhysRevD.95.032002},
while all other intermediate resonances are described by BW functions with masses and widths fixed to the PDG values~\cite{pdg2024}.
The uncertainty related to the background model is estimated by varying the fit range and adopting alternative background parametrizations. 
The uncertainty related to an additional resonance is estimated by varying its treatment. 
For the $K^{0}_{S}K^{0}_{S}\pi^{0}$ mode, the uncertainty includes a contribution from the additional resonance $X(2260)$. 
This inclusion is motivated by the study of the charged channel $J/\psi\rightarrow\gamma K_{S}^{0}K^{\pm}\pi^{\mp}$
\footnote{M. Ablikim \textit{et al}. (BESIII Collaboration), \textit{Observation of the $X(2370)$ in $J/\psi\rightarrow\gamma K_{S}^{0}K^{\pm}\pi^{\mp}$}, publication in preparation.}, 
where the $X(2260)$ is observed in the $K_{S}^{0}K^{\pm}\pi^{\mp}$ mass spectrum with a significance greater than 10$\sigma$, and its fitted mass and width are $2256 \pm 3({\rm stat})~\MeV/c^{2}$ and $74 \pm 10({\rm stat})~\MeV$, respectively. 
The significance of the $X(2260)$ in the $K^{0}_{S}K^{0}_{S}\pi^{0}$ mass spectrum is 2.3$\sigma$, where the low significance is due to the available statistics being more than ten times smaller here than in the charged channel. 
The uncertainty related to the impact of the $X(2260)$ is estimated by incorporating its contribution to the $K^{0}_{S}K^{0}_{S}\pi^{0}$ mass spectrum fit, 
and the mass and width of the $X(2260)$ are fixed to the values measured in the $K_{S}^{0}K^{\pm}\pi^{\mp}$ mass spectrum. 
For the $\pi^{0}\pi^{0}\eta$ mode, a potential additional resonance can be added in the mass spectrum fit with a statistical significance of 4$\sigma$, and its mass and width are fitted to be $2142 \pm 7({\rm stat})~\MeV/c^{2}$ and $186 \pm 48({\rm stat})~\MeV$, respectively. The uncertainty related to the impact of this resonance is estimated by incorporating its contribution to the $\pi^{0}\pi^{0}\eta$ mass spectrum fit. 
For the $a_{0}(980)^{0}\pi^{0}$ mode, since no evidence of the additional resonance is observed, the uncertainty related to the additional resonance is not included. 
All sources of systematic uncertainties are summarized in Table~\ref{tab:sys}. 
\vspace{-4.0mm}

\begin{widetext}

\begin{table*}[tb]
\centering
\renewcommand\arraystretch{1.2}
\setlength{\tabcolsep}{2.5mm}
\caption{Systematic uncertainties on the mass (in $\MeV/c^{2}$), width (in $\MeV$), and significance (in $\sigma$) of the $X(2370)$ in different processes, denoted by $\Delta M$, $\Delta \Gamma$, and $N_{\sigma}$, respectively.}
\vspace{-2.0mm}    
\begin{tabular}{cccccccccc}
\hline\hline
\multicolumn{1}{c}{\multirow{2}{*}{Source}} & \multicolumn{3}{c}{$J/\psi\to\gamma K^{0}_{S}K^{0}_{S}\pi^{0}$} &\multicolumn{3}{c}{$J/\psi\to\gamma \pi^{0}\pi^{0}\eta$} &\multicolumn{3}{c}{\makecell[c]{$J/\psi\to\gamma \pi^{0}\pi^{0}\eta$ \\ ($a_{0}(980)^{0}$ signal region)}}  \\
    &$\Delta M$ &$\Delta \Gamma$ & $N_{\sigma}$  &$\Delta M$ &$\Delta \Gamma$ & $N_{\sigma}$ &$\Delta M$ &$\Delta \Gamma$ & $N_{\sigma}$  \\
    \hline
PHSP parametrization            &   $\pm7$   &$\pm2$  & $>18$ & $\pm3$    & $\pm1$    & $>20$ & $\pm2$   & $\pm3$   & $>12$\\
    \hline
Background model        &   $\pm10$   &$\pm37$  & $>14$ & $\pm9$   & $\pm24$   & $>20$ & $\pm9$   & $\pm58$   & $>9$\\
    \hline
Additional resonance   &   $\pm22$  &$\pm62$  & $>14$  & $\pm4$   &  $\pm56$  & $>20$ & ---   &  ---  & --- \\
    \hline 
    Total               &   $\pm25$  &$\pm72$  & $>14$ & $\pm10$   & $\pm61$   & $>20$ & $\pm9$   & $\pm58$   & $>9$\\

    \hline\hline
\end{tabular}
\label{tab:sys}
\end{table*}

\begin{figure*}[tb]
    \centering
\vspace{-4.0mm}

\subfloat{
  \includegraphics[width=0.45\textwidth]{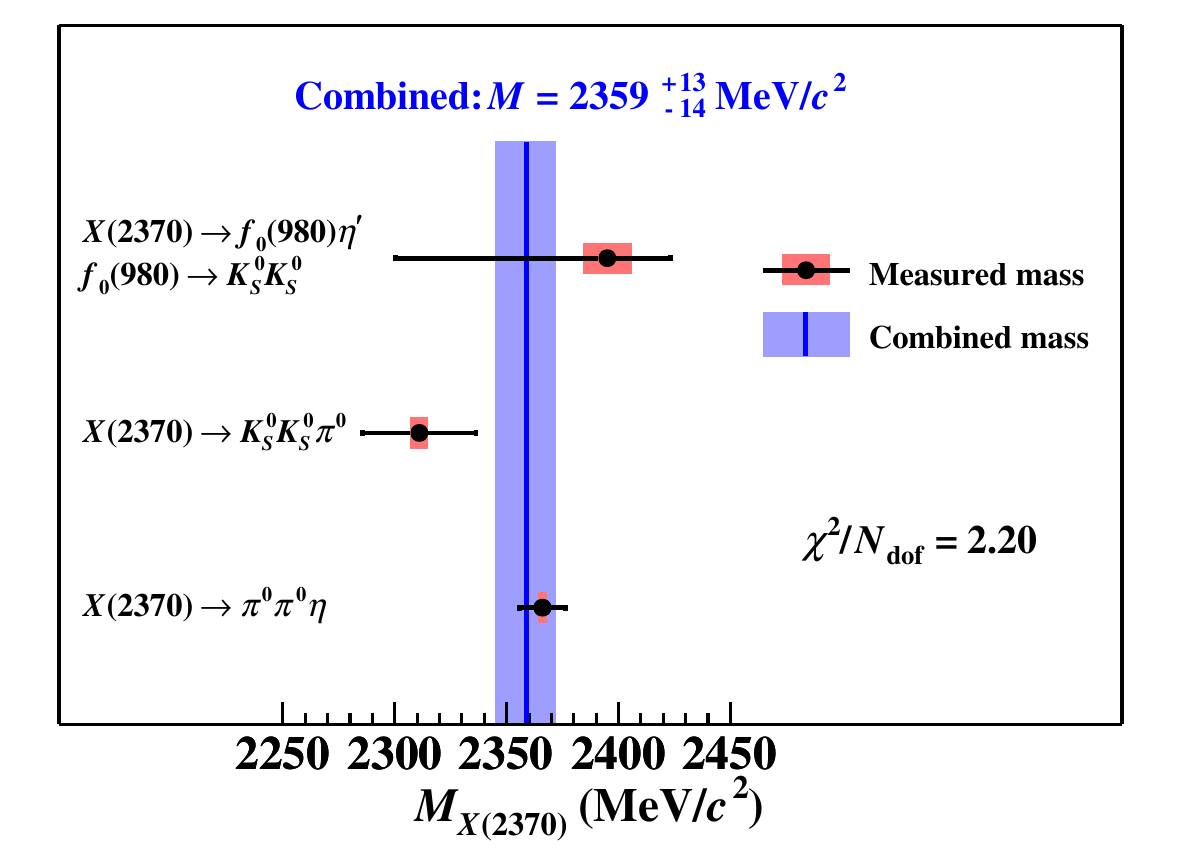}
  \label{fig:Subfigure11}
\put(-215,145){(a)}
}
\hspace{-5.0mm}
\subfloat{
  \includegraphics[width=0.45\textwidth]{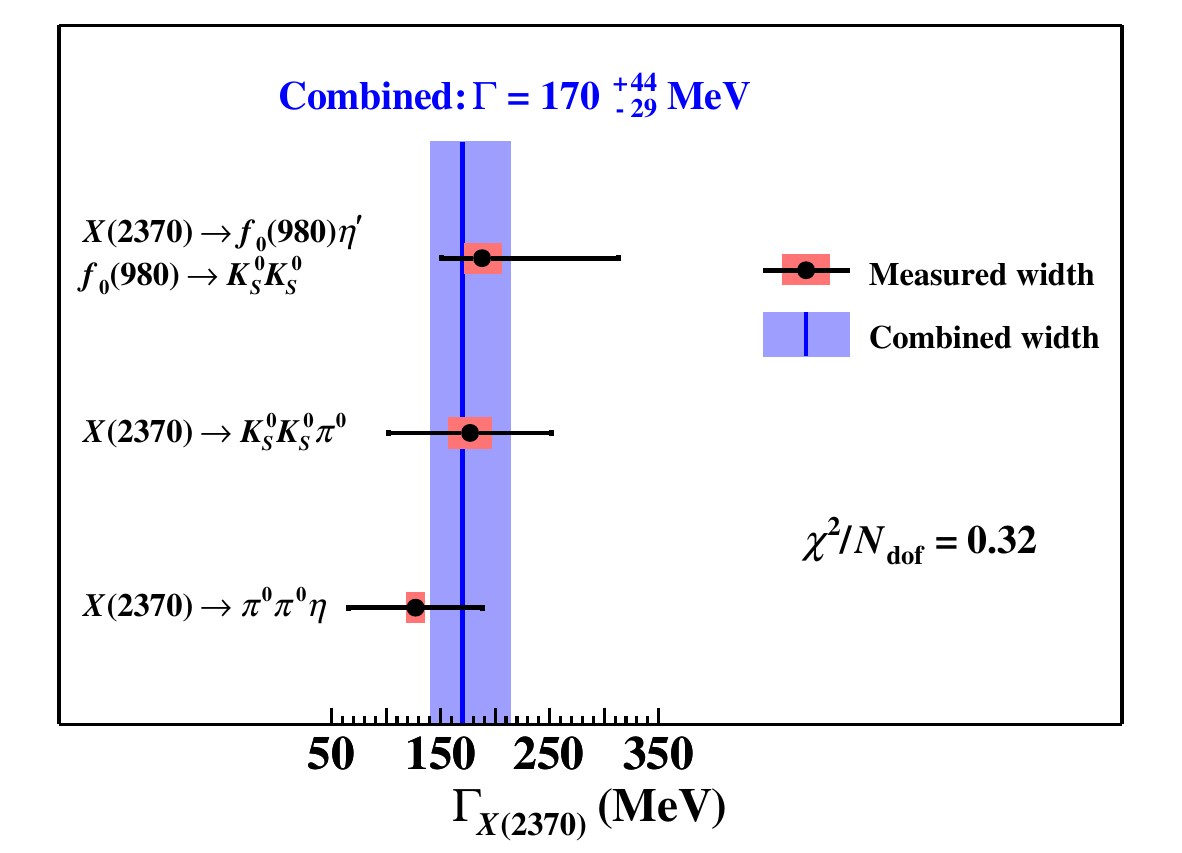}
  \label{fig:Subfigure12}
\put(-215,145){(b)}
}
\vspace{-3.0mm}
\caption{
Combined (a) mass and (b) width of the $X(2370)$. The bars on each black dot indicate the total uncertainty; the horizontal red shaded bands indicate the statistical uncertainties; and the vertical blue lines and blue shaded bands indicate the combined results and their total uncertainties.
}
\label{fig:combination}
\end{figure*}

\end{widetext}

To further improve the measurement precision, the $X(2370)$ masses and widths measured in different decay modes are combined following the procedure recommended by the PDG~\cite{pdg2024}. All $X(2370)$ measurements included in our combination are based on the data sample of 10 billion $J/\psi$ events collected with the BESIII detector, including the studies of the $J/\psi\to\gamma K^{0}_{S}K^{0}_{S}\pi^{0}$ and $J/\psi\to\gamma \pi^{0}\pi^{0}\eta$ processes presented in this paper, as well as the previously reported results for the $J/\psi \to \gamma K^{0}_{S}K^{0}_{S}\eta^{\prime}$ process~\cite{2370jpc}. Considering the statistical and systematic uncertainties, the combined mass and width of the $X(2370)$ are $2359^{+13}_{-14}~\MeV/c^{2}$ and $170^{+44}_{-29}~\MeV$, respectively. The uncertainty of the combined mass is scaled by a factor of 1.48 because $\chi^2/N_{\rm dof} = 2.20$, following the procedure recommended by the PDG~\cite{pdg2024}. Figure~\ref{fig:combination} presents a comparison of the $X(2370)$ resonance parameters measured in the individual channels and the combined result discussed in this paper. The resulting $\chi^{2}/N_{\mathrm{dof}}$ values, shown in each figure and defined in the PDG~\cite{pdg2024}, indicate reasonable consistency among the input measurements.

In summary, using a sample of $(10087 \pm 44) \times 10^{6}$ $J/\psi$ events~\cite{number} collected with the BESIII detector, the $J/\psi\rightarrow\gamma K^{0}_{S}K^{0}_{S}\pi^{0}$ and $J/\psi\rightarrow\gamma \pi^{0}\pi^{0}\eta$ processes are investigated to search for the $X(2370)$. The $X(2370)$ is observed in the $K^{0}_{S}K^{0}_{S}\pi^{0}$ and $\pi^{0}\pi^{0}\eta$ invariant mass spectra with statistical significances greater than 14$\sigma$ and 20$\sigma$, respectively. By further selecting the $a_{0}(980)^{0}$ mass region in the latter process, the decay $X(2370)\to a_{0}(980)^{0}\pi^{0}$ with $a_{0}(980)^{0}\to\pi^{0}\eta$ is observed with a statistical significance greater than $9\sigma$. In the above decay modes, the mass and width of the $X(2370)$ are determined to be
\begin{equation*}
\begin{aligned}
1)~ X(2370)  & \to  K^{0}_{S}K^{0}_{S}\pi^{0} : \\
  M_{X(2370)}       & =2311\pm 4(\mathrm{stat})\pm25(\mathrm{syst}) ~\MeV/c^{2}  ~, \\
   \Gamma_{X(2370)} &=177\pm20(\mathrm{stat})\pm72(\mathrm{syst}) ~\MeV   ~; 
\end{aligned}
\end{equation*}
\vspace{-5mm}

\begin{equation*}
\begin{aligned}
2)~ X(2370) & \to \pi^{0}\pi^{0}\eta : \\
  M_{X(2370)}       & = 2366\pm2(\mathrm{stat})\pm10(\mathrm{syst}) ~\MeV/c^{2} ~, \\
   \Gamma_{X(2370)} &=127\pm9(\mathrm{stat})\pm61(\mathrm{syst}) ~\MeV ~; 
\end{aligned}
\end{equation*}
\vspace{-5mm}

\begin{equation*}
\begin{aligned}
3)~ X(2370)& \to a_{0}(980)^{0}\pi^{0} : \\
  M_{X(2370)}      &=2358\pm5(\mathrm{stat})\pm9(\mathrm{syst})~\MeV/c^{2} ~, \\
  \Gamma_{X(2370)} &=180\pm20(\mathrm{stat})\pm58(\mathrm{syst})~\MeV ~.  
\end{aligned}
\end{equation*}
Based on the results of the first two decay modes observed in this paper, together with the previously reported results in $J/\psi\to\gamma K^{0}_{S}K^{0}_{S}\eta^{\prime}$~\cite{2370jpc}, a combined determination of the $X(2370)$ mass and width is performed, yielding 
\begin{equation*}
\begin{aligned}
  M_{X(2370)}      &=2359^{+13}_{-14}~\MeV/c^{2} ~, \\
  \Gamma_{X(2370)} &=170^{+44}_{-29}~\MeV ~.  
\end{aligned}
\end{equation*}
The combined mass of the $X(2370)$ lies within the mass range predicted by LQCD calculations for the lightest pseudoscalar glueball~\cite{LQCD5,Vadacchino:2023vnc}. 
The observed decay modes of $\pi^{+}\pi^{-}\eta^{\prime}$, $K^{0}_{S}K^{0}_{S}\eta^{\prime}$, $K^{0}_{S}K^{0}_{S}\pi^{0}$, and $\pi^{0}\pi^{0}\eta$, including $a_{0}(980)^{0}\pi^{0}$, show similarities between the $X(2370)$ and $\eta_{c}$ decays.
In addition, 
as shown in the scatter plot of $M_{K^{0}_{S}K^{0}_{S}}$ versus $M_{K^{0}_{S}K^{0}_{S}\eta}$ from a previous study of the $J/\psi \to \gamma K^{0}_{S}K^{0}_{S}\eta$ channel~\cite{1835jpc},
similar behavior is observed in the $K^{0}_{S}K^{0}_{S}\eta$ decay mode. The event yields in both the $X(2370)$ and $\eta_{c}$ mass regions are enhanced in the $f_{0}(1500)$ and $f_{0}(1710)$ mass regions, whereas they are suppressed in the $f_{0}(980)$ mass region.
The properties of the $X(2370)$\textemdash decay pattern similarities to that of $\eta_{c}$, are consistent with those of a pseudoscalar glueball.
To further identify the nature of the $X(2370)$, more studies are needed both experimentally and theoretically.

The BESIII Collaboration thanks the staff of BEPCII (https://cstr.cn/31109.02.BEPC) and the IHEP computing center for their strong support. This work is supported in part by the National Key R\&D Program of China under Contracts Nos. 2025YFA1613900, 2023YFA1606000, 2023YFA1606704; National Natural Science Foundation of China (NSFC) under Contracts Nos. 11635010, 11935015, 11935016, 11935018, 12025502, 12035009, 12035013, 12061131003, 12192260, 12192261, 12192262, 12192263, 12192264, 12192265, 12221005, 12225509, 12235017, 12342502, 12361141819, 12535005; the Chinese Academy of Sciences (CAS) Large-Scale Scientific Facility Program; the Strategic Priority Research Program of Chinese Academy of Sciences under Contract No. XDA0480600; CAS under Contract No. YSBR-101; 100 Talents Program of CAS; China Postdoctoral Science Foundation under Contracts Nos. GZC20252256, 2025M773423; The Institute of Nuclear and Particle Physics (INPAC) and Shanghai Key Laboratory for Particle Physics and Cosmology; Agencia Nacional de Investigación y Desarrollo de Chile (ANID), Chile under Contract No. ANID CCTVal CIA250027; ERC under Contract No. 758462; German Research Foundation DFG under Contract No. FOR5327; Istituto Nazionale di Fisica Nucleare, Italy; Knut and Alice Wallenberg Foundation under Contracts Nos. 2021.0174, 2021.0299, 2023.0315; Ministry of Development of Turkey under Contract No. DPT2006K-120470; National Research Foundation of Korea under Contract No. NRF-2022R1A2C1092335; National Science and Technology fund of Mongolia; Polish National Science Centre under Contract No. 2024/53/B/ST2/00975; STFC (United Kingdom); Swedish Research Council under Contract No. 2019.04595; U. S. Department of Energy under Contract No. DE-FG02-05ER41374. \bibliographystyle{apsrev4-2}
\DIFaddend \bibliography{draft}
\end{document}